\pdfoutput=1

\documentclass[aps,prl,notitlepage,twoside,twocolumn,preprintnumbers,showpacs,showkeys,byrevtex,floatfix,nofootinbib,amsmath,amssymb,]{revtex4-1}
\usepackage{amscd,array,dsfont,graphicx,mathtools,multirow,rotating,tensor,verbatim,youngtab}
\usepackage[bookmarksnumbered,bookmarksopen=true,bookmarksopenlevel=1,linktocpage,pdfstartview=FitH]{hyperref}
\usepackage[all]{hypcap}

\allowdisplaybreaks[1]

\newcolumntype{M}[1]{>{$}{#1}<{$}}

\newcolumntype{M}[1]{>{$}{#1}<{$}}

\newcommand{\sst}[1]{{\scriptscriptstyle #1}}

\def\0{{\sst{(0)}}}
\def\1{{\sst{(1)}}}
\def\2{{\sst{(2)}}}
\def\3{{\sst{(3)}}}
\def\4{{\sst{(4)}}}
\def\5{{\sst{(5)}}}
\def\6{{\sst{(6)}}}
\def\7{{\sst{(7)}}}

\newcommand{\be}{\begin{equation}}
\newcommand{\ee}{\end{equation}}
\def\ba{\begin{array}}
\def\ea{\end{array}}

\newcommand{\bea}{\begin{eqnarray}}
\newcommand{\eea}{\end{eqnarray}}

\DeclareMathOperator{\Hom}{Hom}
\DeclareMathOperator{\SO}{SO}
\DeclareMathOperator{\USp}{USp}
\DeclareMathOperator{\SL}{SL}
\DeclareMathOperator{\SU}{SU}

\newcommand{\alg}{\mathds{A}}
\newcommand{\mf}{\mathfrak}

\newcommand{\R}{\mathds{R}}
\newcommand{\C}{\mathds{C}}
\newcommand{\Q}{\mathds{H}}
\newcommand{\Oct}{\mathds{O}}

\begin{document}

\title{A  magic square from  Yang-Mills squared}
\author{L. Borsten}
\email[]{leron.borsten@imperial.ac.uk}
\author{M. J. Duff}
\email[]{m.duff@imperial.ac.uk}
\author{L. J. Hughes}
\email[]{leo.hughes07@imperial.ac.uk}
\author{S. Nagy}
\email[]{s.nagy11@imperial.ac.uk}

\affiliation{Theoretical Physics, Blackett Laboratory, Imperial College London, London SW7 2AZ, United Kingdom}

\date{\today}

\begin{abstract}

We give a unified description of $D = 3$ super-Yang-Mills theory with $\mathcal{N}=1,2,4,$ and $8$ supersymmeties in terms of the four division algebras: reals ($\R$), complexes ($\C$), quaternions ($\Q$) and octonions ($\Oct$). Tensoring left and right super-Yang-Mills multiplets with $\mathcal{N}=1,2,4,8$ we obtain a magic square   $\R\R$, $\C\R$, $\C\C$, $\Q\R$, $\Q\C$, $\Q\Q$, $\Oct\R$, $\Oct\C$, $\Oct\Q$, $\Oct\Oct$ description of $D=3$ supergravity with $\mathcal{N}=2,3,4,5,6,8,9,10,12,16$. 
\end{abstract}

\pacs{04.65.+e, 02.20.Sv, 11.30.-j}

\keywords{U-duality, magic square, Yang-Mills squared}

\preprint{Imperial/TP/2013/mjd/01}

\maketitle

\paragraph{Introduction} The octonions occupy a privileged position as the largest of the division algebras $\alg$: reals $\R$, complexes $\C$, quaternions $\Q$ and octonions $\Oct$. They provide an intuitive basis for the exceptional Lie groups. For example, the smallest exceptional group $G_2$ preserves the octonionic product. Efforts to understand the remaining exceptional groups geometrically in terms of octonions
 resulted in the Freudenthal-Rozenfeld-Tits magic square \cite{Freudenthal:1954, Freudenthal:1959, Freudenthal:1964, Rosenfeld:1956,Tits:1966,Vinberg:1966} presented in \autoref{tab:ms1}\footnote{There are a variety of magic squares in which different real forms appear. See \cite{Cacciatori:2012cb}  for a comprehensive account in the  context of supergravity. Famously, the $\C, \Q,$ and $\Oct$ rows of one example describe the U-dualities of the aptly named ``magic''  supergravities in $D=5,4,3$ respectively \cite{Gunaydin:1983bi, Gunaydin:1983rk}. In this paper we instead demonstrate the novel appearance of the magic square of \autoref{tab:ms1} in conventional $D=3$ supergravities.}.
 Despite much effort, however, it is fair to say that the ultimate physical significance of octonions and the magic square remains an enigma.
 \begin{table}[ht]
\begin{ruledtabular}
\begin{tabular}{c|ccccc}
 $\alg_L/\alg_R$ & $\R$ & $\C$  & $\Q$  & $\Oct$  \\
 \hline
   $\R$ & $\SL(2, \R)$ & $\SU(2,1)$   & $\USp(4,2)$   & $F_{4(-20)}$   \\
  $\C$ & $\SU(2,1)$ & $\SU(2,1)\times  \SU(2,1)$   & $\SU(4,2)$   & $E_{6(-14)}$   \\
  $\Q$ & $\USp(4,2)$ & $\SU(4,2)$   & $\SO(8,4)$   & $E_{7(-5)}$   \\
   $\Oct$ & $F_{4(-20)}$ & $E_{6(-14)}$   & $E_{7(-5)}$   & $E_{8(8)}$   \\
\end{tabular}
\caption[Magic square of required real forms.]{Magic square \label{tab:ms1}}
\end{ruledtabular}
\end{table}

In apparently different developments, a recurring theme in attempts to understand the quantum theory of gravity is the idea of ``Gravity as the square of Yang-Mills''. This idea of tensoring left $(L)$ and right $(R)$ multiplets appears in many different (but sometimes overlapping) guises: KLT relations in string theory \cite{Kawai:1985xq},  $D=10$ dimensional Type IIA and IIB  supergravity (SG) multiplets   from $D=10$ super Yang-Mills (SYM) multiplets \cite{Green:1987sp}, asymmetric orbifold contructions \cite{Sen:1995ff}, gravity anomalies from gauge anomalies \cite{Antoniadis:1992sa}, (super)gravity   scattering amplitudes from those of (super) Yang-Mills  \cite{Bern:2008qj, Bern:2010ue,Huang:2012wr} in various dimensions etc.

These many-faceted relations have furthered our understanding  of  (super)gravity itself. For example,  the Bern-Carrasco-Johansson (BCJ) color-kinematic duality \cite{Bern:2008qj, Bern:2010ue} has facilitated the computation of higher-loop $D=4, \mathcal{N}=8$ supergravity amplitudes previously regarded as beyond reach. See, for example, \cite{Bern:2009kd} and the references therein.  This  promises to answer the long-standing questions \cite{Duff:1982yw} of if when and how perturbative $\mathcal{N}=8$ supergravity diverges. 

In spite of these remarkable developments, it is still not entirely clear what precisely it means to say gravity is the square of Yang-Mills. For example,  in the supersymmetric context it is not difficult to see that the amount of supersymmetry is given by
\be
[\mathcal{N}_{L}\;{\text{SYM}}] \otimes[\mathcal{N}_{R}\;{\text{SYM}}]\rightarrow[\mathcal{N}=\mathcal{N}_{L}+\mathcal{N}_{R}\;{\text{SG}}],
\ee
but it is harder to see how the other gravitational symmetries arise from those of Yang-Mills.  In particular, supergravities are characterized by non-compact global symmetries $G$ (the so-called U-dualities)  with local compact subgroups $H$, for example $G=E_{7(7)}$ and $H=\SU(8)$ for $\mathcal{N}=8$ supergravity in $D=4$;  whereas the Yang-Mills we start with has global R-symmetries, for example  $\text{R}=\SU(4)$ for $\mathcal{N}=4$ in $D=4$. See \cite{Bianchi:2008pu} for an approach linking $\SU(4)$ to $\SU(8)$ based on scattering amplitudes.

In the present paper we focus on this question in three spacetime dimensions and, in doing so, reveal a magic square of $D=3$ supergravity theories. Hence, looking through the prism of ``gravity $=$ gauge $\times$ gauge'' we  uncover novel structural features of the symmetries in $D=3$ supergravity. Questions of perturbative quantum gravity aside,  understanding supergravity and its symmetries is essential in the context of  string/M-theory, since it constitutes their low-energy effective field theory limit.   In particular, supergravity has been central in exposing the  non-perturbative aspects of string theory. Here, symmetries, especially U-duality, have played a crucial role, for example in constructing black hole solutions, highlighting their significance. 

 In three-dimensional supergravity the  dynamical bosonic degrees of freedom are unified in a $G/H$ coset and, in this sense,  $D=3$ is rather special, throwing light on  the higher-dimensional theories to which it is related by dimensional reduction.  This is particularly true in the present context when generalising the magic square to $D=4, 6, 10$ \cite{Anastasiou:2013hba}. Moreover, $D=3$ is also intrinsically interesting for a number of reasons \cite{Deser:1981wh, Marcus:1983hb, deWit:1992up, Huang:2012wr}, one important example being the remarkable observation that pure three-dimensional quantum gravity is actually solvable \cite{Witten:1988hc, Ashtekar:1989qd}.

Here, we first give a division algebra  $\R,\C,\Q,\Oct$ description of $D=3$ Yang-Mills with $\mathcal{N}=1,2,4,8$, which is of interest in its own right. More remarkable, however, is that tensoring left and right multiplets yields a magic square  $\R\R$, $\C\R$, $\C\C$, $\Q\R$, $\Q\C$, $\Q\Q$, $\Oct\R$, $\Oct\C$, $\Oct\Q$, $\Oct\Oct$ description of $D=3$ supergravity with $\mathcal{N}=2,3,4,5,6,8,9,10,12,16$, as presented in \autoref{tab:3Dsugra}.
For $\mathcal{N}> 8$ the multiplets are those of pure supergravity; for  $\mathcal{N} \leq8$ supergravity is coupled to matter. In both cases the field content  is such that the U-dualities exactly match the groups of \autoref{tab:ms1}.

\begin{table*}
\scriptsize
\begin{ruledtabular}
 \begin{tabular*}{\textwidth}{l|llllllllllll}
&$\R$&$\C$&$\Q$&$\Oct$\\
\hline
&$\mathcal{N}=2, f=4$&$\mathcal{N}=3, f=8$&$\mathcal{N}=5, f=16$&$\mathcal{N}=9, f=32$\\
$\R$&$G=\SL(2,\R)$, $\dim 3$&$G=\SU(2,1)$, $\dim 8$&$G=\USp(4,2)$, $\dim 21$&$G=F_{4(-20)}$, $\dim 52$\\
&$H=\SO(2)$, $\dim 1$&$H=\SU(2) \times \SO(2)$, $\dim 4$&$H=\USp(4)\times \USp(2)$, $\dim13$&$H=\SO(9)$, $\dim 36$\\

&&&&\\

&$\mathcal{N}=3, f=8$&$\mathcal{N}=4, f=16$&$\mathcal{N}=6, f=32$&$\mathcal{N}=10, f=64$\\
$\C$&$G=\SU(2,1)$, $\dim 8$&$G=\SU(2,1)^2$, $\dim 16$&$G=\SU(4,2)$, $\dim 35$&$G=E_{6(-14)}$, $\dim 78$\\
&$H=\SU(2) \times \SO(2)$, $\dim 4$&$H=\SU(2)^2 \times \SO(2)^2$, $\dim 8$&$H=\SU(4) \times \SU(2)\times \SO(2)$, $\dim19$&$H=\SO(10) \times \SO(2)$, $\dim 46$\\

&&&\\

&$\mathcal{N}=5, f=16$&$\mathcal{N}=6, f=32$&$\mathcal{N}=8, f=64$&$\mathcal{N}=12, f=128$\\
$\Q$&$G=\USp(4,2)$, $\dim 21$&$G=\SU(4,2)$, $\dim 35$&$G=\SO(8,4)$, $\dim 66$&$G=E_{7(-5)}$, $\dim 133$\\
&$H=\USp(4)\times \USp(2)$, $\dim13$&$H=\SU(4) \times \SU(2)\times \SO(2)$, $\dim19$&$H=\SO(8)\times \SO(4)$, $\dim34$&$H=\SO(12)\times \SO(3)$, $\dim69$\\

&&&&\\

&$\mathcal{N}=9, f=32$&$\mathcal{N}=10, f=64$&$\mathcal{N}=12, f=128$&$\mathcal{N}=16, f=256$\\
$\Oct$&$G=F_{4(-20)}$, $\dim 52$&$G=E_{6(-14)}$, $\dim 78$&$G=E_{7(-5)}$, $\dim 133$&$G=E_{8(8)}$, $\dim 248$\\
&$H=\SO(9)$, $\dim 36$&$H=\SO(10) \times \SO(2)$, $\dim 46$&$H=\SO(12)\times \SO(3)$, $\dim69$&$H=\SO(16)$, $\dim 120$\\
\end{tabular*}
\caption{Magic square of $D=3$ supergravity theories.  The first row of each entry indicates the amount of supersymmetry $\mathcal{N}$ and the total number of degrees of freedom $f$. The second (third) row indicates the U-duality group $G$ (the maximal compact subgroup $H\subset G$) and its dimension. The scalar fields in each case parametrise the coset $G/H$, where $\dim_\R(G/H)=f/2$. }.
\label{tab:3Dsugra}
\end{ruledtabular}
\end{table*}

Thus not only do $D=3$ supergravities fill out a magic square but their field content and, hence, symmetries are \emph{derived} from squaring Yang-Mills.

\paragraph{Magic square} Magic squares are based on the four division algebras, $\R, \C, \Q$ and $\Oct$, which are of dimension $1,2,4$ and $8$, respectively\footnote{One can also use their split (non-division) cousins to obtain different real forms. See \cite{Barton:2003, Cacciatori:2012cb} and the reference therein for details.}. They can be built, one-by-one, using the Cayley-Dickson doubling procedure starting with $\R$. The reals are ordered, commutative and associative. With each doubling one such property is lost: $\C$ is commutative and associative,  $\Q$ is associative, $\Oct$ is \emph{non-associative}.

An element $x\in\Oct$ may be written $x=x^ae_a$, where $a=0,\ldots,7$,  $x^a\in \R$ and $\{e_a\}$ is a basis with one real $e_0=1$ and  seven $e_i, i=1,\ldots, 7,$ imaginary elements. The octonionic conjugation is denoted by $e_{a}^{*}$, where $e^{*}_{0}=e_0$ and $e^{*}_{i}=-e_i$.
The octonionic multiplication rule is,
\be
e_ae_b=\left(\delta_{a0}\delta_{bc}+\delta_{0b}\delta_{ac}-\delta_{ab}\delta_{0c}+C_{abc}\right)e_c,
\ee
where $C_{abc}$ is totally antisymmetric such that $C_{0bc}=0$.
The non-zero $C_{ijk}$  are given by the Fano plane, see \cite{Baez:2001dm}.

A natural inner product on $\alg$ is defined by
\be
\langle x|y\rangle:=\frac{1}{2}(x\overline{y}+y\overline{x})=x^ay^b\delta_{ab}.\ee

To understand the symmetries of the magic square and its relation to  SYM  we shall need in particular two algebras defined on $\alg$. First, the norm-preserving algebra, 
\be
\mathfrak{so}(\alg):=\{D\in \Hom_\R(\alg) | \langle Dx|y\rangle +\langle x| Dy\rangle=0\},\ee
isomorphic to $\mathfrak{so}(\dim_\R \alg)$. Second, the \emph{triality} algebra
\be
\mathfrak{tri}(\alg):=\{(A, B, C) | A(xy)=(Bx)y+x(Cy)\}\ee
where $A,B,C\in\mathfrak{so}(\alg)$. For $\alg=\R, \C, \Q, \Oct$ we have $\mf{tri}(\alg)\cong \O, \mf{so}(2)\oplus \mf{so}(2), \mf{so}(3)\oplus \mf{so}(3)\oplus \mf{so}(3), \mf{so}(8)$ \cite{Barton:2003}.

The specific magic square presented in \autoref{tab:ms1} was first obtained in \cite{Cacciatori:2012cb} using a version of the Tits construction  based on a \emph{Lorentzian} Jordan algebra.  The two algebras $\alg_L, \alg_R$ enter this definition on distinct footings; the ``magic'' of the  square is its symmetry under the exchange $\alg_L\leftrightarrow \alg_R$, which is obscured by their undemocratic treatment.

For the purposes of squaring SYM a manifestly $\alg_L\leftrightarrow \alg_R$ symmetric formulation of the square is required. This is achieved by adapting  the \emph{triality algebra} construction introduced by Barton and Sudbery \cite{Barton:2003}.
Our definition\footnote{Note, this is not quite the triality construction as defined in \cite{Barton:2003}. We will not present the details here, but it can be easily obtained by making a slight modification to the commutators in $3(\alg_L\otimes\alg_R)$ w.r.t. those appearing in \cite{Barton:2003}. Here, we have used $\oplus$ and $+$ to distinguish  the direct sum between Lie algebras and vector spaces, i.e. only if $[\mathfrak{g}, \mathfrak{h}]=0$ do we use $\mathfrak{g}\oplus\mathfrak{h}$.
} of \autoref{tab:ms1}  is given by,
\be\label{eq:tri}
\mf{L}_3(\alg_L, \alg_R)\cong\mf{tri}(\alg_L)\oplus \mf{tri}(\alg_R)+3(\alg_L\otimes\alg_R).
\ee

We shall also need a magic square of the maximal compact subalgebras of \autoref{tab:ms1}, given in \autoref{tab:ms3}.
This is given by the \emph{reduced} triality construction,
\be\label{eq:redtri}
\mf{L}_1(\alg_L, \alg_R):=\mf{tri}(\alg_L)\oplus \mf{tri}(\alg_R)+(\alg_L\otimes\alg_R),
\ee
which is easily obtained from \eqref{eq:tri}. 

\begin{table*}
\scriptsize
\begin{ruledtabular}
\begin{tabular*}{\textwidth}{c|ccccc}
 & $\R$ & $\C$  & $\Q$  & $\Oct$  \\
 \hline
 
   $\R$ & $\SO(2)$ & $\SO(3)\times \SO(2)$   & $\SO(5)\times \SO(3)$   & $\SO(9)$   \\
  $\C$ & $\SO(3)\times \SO(2)$ & $[\SO(3)\times \SO(2)]^2$   & $\SO(6)\times \SO(3)\times \SO(2)$   & $\SO(10)\times \SO(2)$   \\
  $\Q$ & $\SO(5)\times \SO(3)$ & $\SO(6)\times \SO(3)\times \SO(2)$   &$\SO(8)\times \SO(4)$  & $\SO(12)\times \SO(3)$   \\
   $\Oct$ & $\SO(9)$ & $\SO(10)\times \SO(2)$   & $\SO(12)\times \SO(3)$   & $\SO(16)$   \\
\end{tabular*}
\caption[Magic square of required real forms.]{Magic square of maximal compact subgroups.  \label{tab:ms3}}
\end{ruledtabular}
\end{table*}
\begin{table*}[ht]
\scriptsize
\begin{ruledtabular}
\begin{tabular*}{\textwidth}{c|ccccc}
 $\alg_L/\alg_R$ & ${A}_\mu(R) \in \text{Re}\alg_R$ & ${\phi}(R) \in \text{Im}\alg_R$ &${\lambda}(R)\in\alg_R  $\\[3pt]
 \hline
 &\\[-6pt]
   $A_\mu(L) \in \text{Re}\alg_L$ & $g_{\mu\nu}+\varphi\in \text{Re}\alg_L\otimes \text{Re}\alg_R$ & $\varphi\in \text{Re}\alg_L\otimes \text{Im}\alg_R$   & $\Psi_\mu+\chi \in \text{Re}\alg_L\otimes \alg_R$     \\[5pt]
  $\phi(L) \in \text{Im}\alg_L$ & $\varphi\in \text{Im}\alg_L\otimes \text{Re}\alg_R$ & $\varphi\in \text{Im}\alg_L\otimes \text{Im}\alg_R$   & $\chi \in \text{Im}\alg_L\otimes \alg_R$     \\[3pt]
  $\lambda(L)\in\alg_L$ & $\Psi_\mu+\chi \in \alg_L\otimes \text{Re}\alg_R$  & $\chi \in \alg_L\otimes \text{Im}\alg_R$   &$\varphi\in \alg_L\otimes \alg_R$     \\[3pt]
\end{tabular*}
\caption[Magic square of required real forms.]{Tensor product of left/right $(\alg_L/\alg_R)$ SYM multiplets, using $\SO(1,2)$ spacetime    reps and dualising all $p$-forms. \label{tab:tensor}}
\end{ruledtabular}
\end{table*}
\paragraph{ $\R, \C, \Q, \Oct$ description of $D=3, \mathcal{N}=1,2,4,8$  Yang-Mills} The    $D=3$, $\mathcal{N}=8$ SYM Lagrangian is given by
\be
\begin{split}
\mathcal{L}=&-\tfrac{1}{4}F^A_{\mu\nu}F^{A\mu\nu}-\tfrac{1}{2}D_\mu\phi_i^{A}D^\mu\phi_i^A+i\bar{\lambda}_a^{ A}\gamma^\mu D_\mu\lambda_a^{A}  \\ 
& -\tfrac{1}{4}g^2f_{BC}{}^Af_{DE}{}^A\phi_i^B\phi_i^D\phi^C_j\phi^E_j  \\&-gf_{BC}{}^A\phi^B_i \bar{\lambda}^{ Aa}\Gamma^i_{ab} \lambda^{Cb},
\end{split}
\ee
where $\Gamma^i_{ab}$, $i=1,\ldots,7$, $a,b=0,\ldots,7$, belongs to the SO(7) Clifford algebra. The key observation is that this gamma matrix can be represented by the octonionic structure constants,
\be
\Gamma^i_{ab}=i(\delta_{bi}\delta_{a0}-\delta_{b0}\delta_{ai}+C_{iab}),
\ee
which allows us to rewrite the action over octonionic fields. If we replace $\Oct$ with a general division algebra $\alg$, the result is $\mathcal{N}=1,2,4,8$ over $\R,\C,\Q,\Oct$:
\be
\begin{split}
\mathcal{L}=&- \tfrac{1}{4}F^A_{\mu\nu}F^{A\mu\nu}-\tfrac{1}{2}D_\mu\phi^{*A}D^\mu\phi^A+i\bar{\lambda}^{ A}\gamma^\mu D_\mu\lambda^A  \\ 
&-\tfrac{1}{4}g^2f_{BC}{}^Af_{DE}{}^A\langle\phi^B|\phi^D\rangle\langle\phi^C|\phi^E\rangle \hspace{.95cm}  \\
 &+\tfrac{i}{2}gf_{BC}{}^A\left((\bar{\lambda}^{A}\phi^B)\lambda^C-\bar{\lambda}^{A}(\phi^{*B}\lambda^C)\right),
\end{split}
\ee
where $\phi=\phi^i e_i$ is an Im$\mathds{A}$-valued scalar field, $\lambda=\lambda^a e_a$ is an $\mathds{A}$-valued two-component spinor and $\bar{\lambda}=\bar{\lambda}^ae_a^*$. Note, since $\lambda^a$ is anti-commuting we are dealing with the algebra of octonions defined over the Grassmanns.

The supersymmetry transformations in this language are given by
\begin{eqnarray}
\delta\lambda^A&&=\frac{1}{2}(F^{A\mu\nu}+\varepsilon^{\mu\nu\rho}D_\rho\phi^A)\sigma_{\mu\nu}\epsilon-\frac{1}{4}gf_{BC}{}^A\phi^B(\phi^C\epsilon), \nonumber\\
\delta A_\mu^A&&=\frac{i}{2}(\bar{\epsilon}\gamma_\mu\lambda^A-\bar{\lambda}^{ A}\gamma_\mu\epsilon),\\
\delta \phi^A&&= \frac{i}{2}e_i[(\bar{\epsilon} e_i)\lambda^A-\bar{\lambda}^{ A}(e_i\epsilon)],\nonumber
\end{eqnarray}
where $\epsilon$ is an $\mathds{A}$-valued two-component spinor and $\sigma_{\mu\nu}$ are the generators of $\SL(2,\R)\cong\SO(1,2)$. We note that the final term in the transformation of $\lambda^A$ can be rewritten as
\be
-\frac{1}{4}gf_{BC}{}^A\phi^B(\phi^C\epsilon)=\frac{1}{2}gf_{BC}{}^A\phi^B_i\phi^C_j e_a\Gamma^{ij}_{ab}\epsilon_b,
\ee
where the $\Gamma^{ij}$ are the generators of $\SO(n-1)$ in the spinor representation, illustrating the close relationship between division algebraic multiplication and Clifford algebras \cite{Baez:2001dm}.

The form of the first term in the $\lambda^A$ transformation also highlights the vector's status as the missing real part of the Im$\mathds{A}$-valued scalar field. Indeed, in the free $g=0$ theory one may dualise the vector to a scalar to obtain a full $\mathds{A}$-valued field.

\paragraph{Squaring Yang-Mills} Having cast the  magic square  in terms of a manifestly $\alg_L\leftrightarrow\alg_R$ symmetric triality algebra construction, and having written $\mathcal{N}=1,2,4,8$ SYM in terms of fields valued in $\R,\C,\Q,\Oct$ we shall now obtain  the  magic square of supergravities in \autoref{tab:3Dsugra}, with symmetries $G$ (\autoref{tab:ms1}) and $H$ (\autoref{tab:ms3}), by ``squaring'' $\mathcal{N}=1,2,4,8$ SYM.

Taking a left SYM multiplet $\{A_\mu(L) \in \text{Re}\alg_L, \phi(L) \in \text{Im}\alg_L, \lambda(L)\in\alg_L\}$ and tensoring it with a right multiplet $\{{A}_\mu(R) \in \text{Re}\alg_R, {\phi}(R) \in \text{Im}\alg_R,{\lambda}(R)\in\alg_R\}$ we obtain the field content of a supergravity theory valued in both $\alg_L$ and $\alg_R$. See \autoref{tab:tensor}.  Note, the left/right SYM R-symmetries act on each slot of the $\alg_L, \alg_R$ tensor products.

Grouping spacetime fields of the same type we find,
\be\label{eq:sugrafields}
g_{\mu\nu}  \in \R, \quad
\Psi_{\mu} \in  \begin{pmatrix} \alg_L\\ \alg_R\end{pmatrix}, \quad
\varphi, \chi  \in  \begin{pmatrix} \alg_L\otimes \alg_R\\ \alg_L\otimes \alg_R\end{pmatrix}.
\ee
The $\R$-valued graviton and $\alg_L\oplus \alg_R$-valued gravitino carry no degrees of freedom. The $(\alg_L\otimes\alg_R)^2$-valued scalar and Majorana spinor each have $2(\dim \alg_L\times\dim\alg_R)$ degrees of freedom.   

As we have already mentioned, the $\mathcal{N}>8$ supergravities in $D=3$ are unique, all fields belonging to  the gravity multiplet, while those with 
$\mathcal{N}\leq 8$ may be coupled to $k$ additional matter 
multiplets \cite{Marcus:1983hb, deWit:1992up}.  
 The real miracle is that tensoring left and right SYM multiplets yields the field content of  
 $\mathcal{N}=2,3,4,5,6,8$ supergravity with $k=1, 1, 2, 1, 2, 4$: just the right matter content to produce the U-duality groups appearing in \autoref{tab:ms1}.

At this stage we should emphasise that we have thus far neglected the role of the Yang-Mills gauge group when reconstructing the supergravity multiplet. However, the two (possibly distinct) left/right  gauge groups may be accommodated  by introducing a `spectator' scalar field valued in the bi-adjoint of gauge left and gauge right. This field appears as a factor on the gravitational side when  tensoring the left/right SYM fields,  thus accounting for the gauge indices. Interestingly, this  seems to be superficially consistent with the observation, originally made in the context of twistor diagrams in \cite{Hodges:2011wm} and confirmed explicitly solution by solution in \cite{Cachazo:2013iea},  that at tree-level the product of two SYM amplitudes (or, to be precise, their integrands) produces a gravitational amplitude, but with an additional colour factor  that corresponds precisely to the appropriate amplitude for the spectator field with a cubic Lagrangian. Moreover, the scalar/gauge/gravity amplitude formulae of \cite{Cachazo:2013iea} where recently \emph{derived} using  ambitwistor strings  \cite{Mason:2013sva}. From this perspective   ``Yang-Mills $\times$ Yang-Mills $=$ gravity'' is replaced by ``Yang-Mills $\times$ Yang-Mills $=$ gravity $\times$ $\phi^3$'' and  leads to the BCJ color-kinematic duality  \cite{Cachazo:2013iea}. Peeling away the spectator field one can build the supergravity Lagrangian, although the relation to the original SYM fields is muddied in the process. We postpone these important considerations for future work.

Returning to the subject in hand, the largest linearly realised global symmetry of these theories  is $H$, which has Lie algebra given by the reduced triality construction \eqref{eq:redtri}.  Consequently, we expect the fields in \eqref{eq:sugrafields} to carry linear representations of $H$. The metric is a singlet, while $\Psi_\mu, \varphi$ and $\chi$ transform as a vector, spinor and conjugate spinor, respectively. Fortunately, $\alg_L\oplus \alg_R$ and $(\alg_L\otimes\alg_R)^2$ are precisely the representation spaces of the vector and (conjugate) spinor. For example, in the maximal case of $\alg_L, \alg_R=\Oct$, we have the $\mathbf{16}, \mathbf{128}$ and $\mathbf{128'}$ of $\SO(16)$.  The distinction between spinor and conjugate spinor in terms of $(\Oct_L\otimes\Oct_R)^2$ is encoded in the division-algebraic realisation of the Lie algebra action, which is inherited from the left/right SYM. For example, consider $x, y\in\Oct$  transforming respectively as the $\mathbf{8}_s$ and $\mathbf{8}_c$ of $\SO(8)$. A subset of $\SO(8)$ generators  are given by left/right multiplications with  elements $a\in\text{Im}\Oct$ under which    $x\mapsto ax$ implies $y\mapsto ya$. 

The U-dualities $G$ are realised non-linearly on the scalars, which parametrise the symmetric spaces $G/H$.  This can be  understood using the remarkable identity relating the projective planes\footnote{To be precise, these are Lorentzian counterparts of the more familiar compact projective planes, for example $\SO(3)/\SO(2)$ vs. $\SL(2, \R)/\SO(2)$. The Lorentzian Cayley plane may be defined using a Lorentzian Jordan algebra  \cite{Baez:2001dm, Cacciatori:2012cb}.} over $(\alg_L\otimes\alg_R)^2$ to  $G/H$, 
\be
(\alg_L\otimes\alg_R)\mathds{P}^2 \cong G/H.
\ee
The scalar fields may be regarded as points in division-algebraic projective planes. The tangent space at any point of $(\alg_L\otimes\alg_R)\mathds{P}^2$ is just $(\alg_L\otimes\alg_R)^2$, the required representation space  of $H$.
The \emph{Cayley plane} $\Oct\mathds{P}^2$, with isometry group $F_{4(-52)}$, is a classic example: $F_{4(-52)}/\text{Spin}(9)\cong (\R\otimes\Oct)\mathds{P}^2=\Oct\mathds{P}^2$. The tangent space at any point of $\Oct\mathds{P}^2$ is $\Oct^2$,  the spinor of $\text{Spin}(9)$ as required. Note,  the cases  $(\C\otimes\Oct)\mathds{P}^2, (\Q\otimes\Oct)\mathds{P}^2, (\Oct\otimes\Oct)\mathds{P}^2$ are not strictly speaking  projective spaces, but  nevertheless constitute geometries which may be identified with $G/H$ \cite{Baez:2001dm, Freudenthal:1964, Landsberg2001477}.

We conclude by noting that this construction can be extended to $D=\dim(\mathds{A})+2$ by exploiting the $
\mathfrak{so}(1,\dim(\mathds{A})+1)\cong \mathfrak{sl}(2,\mathds{A})$ Lie algebra isomorphisms
 (in the sense of \cite{Sudbery:1984}). This results in a \emph{magic pyramid} of supergravities: the magic square described here forms the base in $D=3$ while type II supergravity sits at the apex in $D=10$. The U-dualities are determined by a \emph{magic pyramid formula}, generalising \eqref{eq:tri}, which is parametrised by three division algebras, one for spacetime and two for left/right SYM \cite{Anastasiou:2013hba}.

\begin{acknowledgements}

We thank David Skinner for bringing the role of the $\phi^3$-theory to our attention. The work of L. B. is supported by an Imperial College Junior Research Fellowship. The work of M. J. D. is supported by the STFC under rolling Grant No. ST/G000743/1. L. J. H. and S. N. are supported by STFC Ph.D. studentships. 

\end{acknowledgements}


\begin{thebibliography}{32}%
\makeatletter
\providecommand \@ifxundefined [1]{%
 \@ifx{#1\undefined}
}%
\providecommand \@ifnum [1]{%
 \ifnum #1\expandafter \@firstoftwo
 \else \expandafter \@secondoftwo
 \fi
}%
\providecommand \@ifx [1]{%
 \ifx #1\expandafter \@firstoftwo
 \else \expandafter \@secondoftwo
 \fi
}%
\providecommand \natexlab [1]{#1}%
\providecommand \enquote  [1]{``#1''}%
\providecommand \bibnamefont  [1]{#1}%
\providecommand \bibfnamefont [1]{#1}%
\providecommand \citenamefont [1]{#1}%
\providecommand \href@noop [0]{\@secondoftwo}%
\providecommand \href [0]{\begingroup \@sanitize@url \@href}%
\providecommand \@href[1]{\@@startlink{#1}\@@href}%
\providecommand \@@href[1]{\endgroup#1\@@endlink}%
\providecommand \@sanitize@url [0]{\catcode `\\12\catcode `\$12\catcode
  `\&12\catcode `\#12\catcode `\^12\catcode `\_12\catcode `\%12\relax}%
\providecommand \@@startlink[1]{}%
\providecommand \@@endlink[0]{}%
\providecommand \url  [0]{\begingroup\@sanitize@url \@url }%
\providecommand \@url [1]{\endgroup\@href {#1}{\urlprefix }}%
\providecommand \urlprefix  [0]{URL }%
\providecommand \Eprint [0]{\href }%
\providecommand \doibase [0]{http://dx.doi.org/}%
\providecommand \selectlanguage [0]{\@gobble}%
\providecommand \bibinfo  [0]{\@secondoftwo}%
\providecommand \bibfield  [0]{\@secondoftwo}%
\providecommand \translation [1]{[#1]}%
\providecommand \BibitemOpen [0]{}%
\providecommand \bibitemStop [0]{}%
\providecommand \bibitemNoStop [0]{.\EOS\space}%
\providecommand \EOS [0]{\spacefactor3000\relax}%
\providecommand \BibitemShut  [1]{\csname bibitem#1\endcsname}%
\let\auto@bib@innerbib\@empty
\bibitem [{\citenamefont {Freudenthal}(1954)}]{Freudenthal:1954}%
  \BibitemOpen
  \bibfield  {author} {\bibinfo {author} {\bibfnamefont {H.}~\bibnamefont
  {Freudenthal}},\ }\href@noop {} {\bibfield  {journal} {\bibinfo  {journal}
  {Nederl. Akad. Wetensch. Proc. Ser.}\ }\textbf {\bibinfo {volume} {57}},\
  \bibinfo {pages} {218} (\bibinfo {year} {1954})}\BibitemShut {NoStop}%
\bibitem [{\citenamefont {Freudenthal}(1959)}]{Freudenthal:1959}%
  \BibitemOpen
  \bibfield  {author} {\bibinfo {author} {\bibfnamefont {H.}~\bibnamefont
  {Freudenthal}},\ }\href@noop {} {\bibfield  {journal} {\bibinfo  {journal}
  {Nederl. Akad. Wetensch. Proc. Ser.}\ }\textbf {\bibinfo {volume} {A62}},\
  \bibinfo {pages} {466} (\bibinfo {year} {1959})}\BibitemShut {NoStop}%
\bibitem [{\citenamefont {Freudenthal}(1964)}]{Freudenthal:1964}%
  \BibitemOpen
  \bibfield  {author} {\bibinfo {author} {\bibfnamefont {H.}~\bibnamefont
  {Freudenthal}},\ }\href@noop {} {\bibfield  {journal} {\bibinfo  {journal}
  {Adv. Math.}\ }\textbf {\bibinfo {volume} {1}},\ \bibinfo {pages} {145}
  (\bibinfo {year} {1964})}\BibitemShut {NoStop}%
\bibitem [{\citenamefont {Rosenfeld}(1956)}]{Rosenfeld:1956}%
  \BibitemOpen
  \bibfield  {author} {\bibinfo {author} {\bibfnamefont {B.~A.}\ \bibnamefont
  {Rosenfeld}},\ }\href@noop {} {\bibfield  {journal} {\bibinfo  {journal}
  {Dokl. Akad. Nauk. SSSR}\ }\textbf {\bibinfo {volume} {106}},\ \bibinfo
  {pages} {600} (\bibinfo {year} {1956})}\BibitemShut {NoStop}%
\bibitem [{\citenamefont {Tits}(1966)}]{Tits:1966}%
  \BibitemOpen
  \bibfield  {author} {\bibinfo {author} {\bibfnamefont {J.}~\bibnamefont
  {Tits}},\ }\href@noop {} {\bibfield  {journal} {\bibinfo  {journal} {Indag.
  Math.}\ }\textbf {\bibinfo {volume} {28}},\ \bibinfo {pages} {223} (\bibinfo
  {year} {1966})}\BibitemShut {NoStop}%
\bibitem [{\citenamefont {Vinberg}(1966)}]{Vinberg:1966}%
  \BibitemOpen
  \bibfield  {author} {\bibinfo {author} {\bibfnamefont {E.~B.}\ \bibnamefont
  {Vinberg}},\ }\href@noop {} {\bibfield  {journal} {\bibinfo  {journal} {Tr.
  Semin. Vektorn. Tr. Semin. Vektorn. Tensorn. Anal.}\ }\textbf {\bibinfo
  {volume} {13}} (\bibinfo {year} {1966})}\BibitemShut {NoStop}%
\bibitem [{\citenamefont {Cacciatori}\ \emph {et~al.}(2012)\citenamefont
  {Cacciatori}, \citenamefont {Cerchiai},\ and\ \citenamefont
  {Marrani}}]{Cacciatori:2012cb}%
  \BibitemOpen
  \bibfield  {author} {\bibinfo {author} {\bibfnamefont {S.~L.}\ \bibnamefont
  {Cacciatori}}, \bibinfo {author} {\bibfnamefont {B.~L.}\ \bibnamefont
  {Cerchiai}}, \ and\ \bibinfo {author} {\bibfnamefont {A.}~\bibnamefont
  {Marrani}},\ }\href@noop {} {\  (\bibinfo {year} {2012})},\ \Eprint
  {http://arxiv.org/abs/1208.6153} {arXiv:1208.6153 [math-ph]} \BibitemShut
  {NoStop}%
\bibitem [{\citenamefont {G{\"u}naydin}\ \emph {et~al.}(1984)\citenamefont
  {G{\"u}naydin}, \citenamefont {Sierra},\ and\ \citenamefont
  {Townsend}}]{Gunaydin:1983bi}%
  \BibitemOpen
  \bibfield  {author} {\bibinfo {author} {\bibfnamefont {M.}~\bibnamefont
  {G{\"u}naydin}}, \bibinfo {author} {\bibfnamefont {G.}~\bibnamefont
  {Sierra}}, \ and\ \bibinfo {author} {\bibfnamefont {P.~K.}\ \bibnamefont
  {Townsend}},\ }\href {\doibase 10.1016/0550-3213(84)90142-1} {\bibfield
  {journal} {\bibinfo  {journal} {Nucl. Phys.}\ }\textbf {\bibinfo {volume}
  {B242}},\ \bibinfo {pages} {244} (\bibinfo {year} {1984})}\BibitemShut
  {NoStop}%
\bibitem [{\citenamefont {G{\"u}naydin}\ \emph {et~al.}(1983)\citenamefont
  {G{\"u}naydin}, \citenamefont {Sierra},\ and\ \citenamefont
  {Townsend}}]{Gunaydin:1983rk}%
  \BibitemOpen
  \bibfield  {author} {\bibinfo {author} {\bibfnamefont {M.}~\bibnamefont
  {G{\"u}naydin}}, \bibinfo {author} {\bibfnamefont {G.}~\bibnamefont
  {Sierra}}, \ and\ \bibinfo {author} {\bibfnamefont {P.~K.}\ \bibnamefont
  {Townsend}},\ }\href {\doibase 10.1016/0370-2693(83)90108-9} {\bibfield
  {journal} {\bibinfo  {journal} {Phys. Lett.}\ }\textbf {\bibinfo {volume}
  {B133}},\ \bibinfo {pages} {72} (\bibinfo {year} {1983})}\BibitemShut
  {NoStop}%
\bibitem [{\citenamefont {Kawai}\ \emph {et~al.}(1986)\citenamefont {Kawai},
  \citenamefont {Lewellen},\ and\ \citenamefont {Tye}}]{Kawai:1985xq}%
  \BibitemOpen
  \bibfield  {author} {\bibinfo {author} {\bibfnamefont {H.}~\bibnamefont
  {Kawai}}, \bibinfo {author} {\bibfnamefont {D.}~\bibnamefont {Lewellen}}, \
  and\ \bibinfo {author} {\bibfnamefont {S.}~\bibnamefont {Tye}},\ }\href
  {\doibase 10.1016/0550-3213(86)90362-7} {\bibfield  {journal} {\bibinfo
  {journal} {Nucl.Phys.}\ }\textbf {\bibinfo {volume} {B269}},\ \bibinfo
  {pages} {1} (\bibinfo {year} {1986})}\BibitemShut {NoStop}%
\bibitem [{\citenamefont {Green}\ \emph {et~al.}(1987)\citenamefont {Green},
  \citenamefont {Schwarz},\ and\ \citenamefont {Witten}}]{Green:1987sp}%
  \BibitemOpen
  \bibfield  {author} {\bibinfo {author} {\bibfnamefont {M.~B.}\ \bibnamefont
  {Green}}, \bibinfo {author} {\bibfnamefont {J.~H.}\ \bibnamefont {Schwarz}},
  \ and\ \bibinfo {author} {\bibfnamefont {E.}~\bibnamefont {Witten}},\
  }\href@noop {} {\emph {\bibinfo {title} {{Superstring Theory vol. 1:
  Introduction}}}},\ Cambridge Monographs on Mathematical Physics\ (\bibinfo
  {publisher} {Cambridge University Press},\ \bibinfo {address} {Cambridge,
  UK},\ \bibinfo {year} {1987})\ \bibinfo {note} {469 p}\BibitemShut {NoStop}%
\bibitem [{\citenamefont {Sen}\ and\ \citenamefont {Vafa}(1995)}]{Sen:1995ff}%
  \BibitemOpen
  \bibfield  {author} {\bibinfo {author} {\bibfnamefont {A.}~\bibnamefont
  {Sen}}\ and\ \bibinfo {author} {\bibfnamefont {C.}~\bibnamefont {Vafa}},\
  }\href {\doibase 10.1016/0550-3213(95)00498-H} {\bibfield  {journal}
  {\bibinfo  {journal} {Nucl. Phys.}\ }\textbf {\bibinfo {volume} {B455}},\
  \bibinfo {pages} {165} (\bibinfo {year} {1995})},\ \Eprint
  {http://arxiv.org/abs/hep-th/9508064} {arXiv:hep-th/9508064} \BibitemShut
  {NoStop}%
\bibitem [{\citenamefont {Antoniadis}\ \emph {et~al.}(1992)\citenamefont
  {Antoniadis}, \citenamefont {Gava},\ and\ \citenamefont
  {Narain}}]{Antoniadis:1992sa}%
  \BibitemOpen
  \bibfield  {author} {\bibinfo {author} {\bibfnamefont {I.}~\bibnamefont
  {Antoniadis}}, \bibinfo {author} {\bibfnamefont {E.}~\bibnamefont {Gava}}, \
  and\ \bibinfo {author} {\bibfnamefont {K.}~\bibnamefont {Narain}},\ }\href
  {\doibase 10.1016/0370-2693(92)90009-S} {\bibfield  {journal} {\bibinfo
  {journal} {Phys.Lett.}\ }\textbf {\bibinfo {volume} {B283}},\ \bibinfo
  {pages} {209} (\bibinfo {year} {1992})},\ \Eprint
  {http://arxiv.org/abs/hep-th/9203071} {arXiv:hep-th/9203071 [hep-th]}
  \BibitemShut {NoStop}%
\bibitem [{\citenamefont {Bern}\ \emph {et~al.}(2008)\citenamefont {Bern},
  \citenamefont {Carrasco},\ and\ \citenamefont {Johansson}}]{Bern:2008qj}%
  \BibitemOpen
  \bibfield  {author} {\bibinfo {author} {\bibfnamefont {Z.}~\bibnamefont
  {Bern}}, \bibinfo {author} {\bibfnamefont {J.}~\bibnamefont {Carrasco}}, \
  and\ \bibinfo {author} {\bibfnamefont {H.}~\bibnamefont {Johansson}},\ }\href
  {\doibase 10.1103/PhysRevD.78.085011} {\bibfield  {journal} {\bibinfo
  {journal} {Phys.Rev.}\ }\textbf {\bibinfo {volume} {D78}},\ \bibinfo {pages}
  {085011} (\bibinfo {year} {2008})},\ \Eprint {http://arxiv.org/abs/0805.3993}
  {arXiv:0805.3993 [hep-ph]} \BibitemShut {NoStop}%
\bibitem [{\citenamefont {Bern}\ \emph {et~al.}(2010)\citenamefont {Bern},
  \citenamefont {Carrasco},\ and\ \citenamefont {Johansson}}]{Bern:2010ue}%
  \BibitemOpen
  \bibfield  {author} {\bibinfo {author} {\bibfnamefont {Z.}~\bibnamefont
  {Bern}}, \bibinfo {author} {\bibfnamefont {J.~J.~M.}\ \bibnamefont
  {Carrasco}}, \ and\ \bibinfo {author} {\bibfnamefont {H.}~\bibnamefont
  {Johansson}},\ }\href {\doibase 10.1103/PhysRevLett.105.061602} {\bibfield
  {journal} {\bibinfo  {journal} {Phys.Rev.Lett.}\ }\textbf {\bibinfo {volume}
  {105}},\ \bibinfo {pages} {061602} (\bibinfo {year} {2010})},\ \Eprint
  {http://arxiv.org/abs/1004.0476} {arXiv:1004.0476 [hep-th]} \BibitemShut
  {NoStop}%
\bibitem [{\citenamefont {Huang}\ and\ \citenamefont
  {Johansson}(2013)}]{Huang:2012wr}%
  \BibitemOpen
  \bibfield  {author} {\bibinfo {author} {\bibfnamefont {Y.-t.}\ \bibnamefont
  {Huang}}\ and\ \bibinfo {author} {\bibfnamefont {H.}~\bibnamefont
  {Johansson}},\ }\href {\doibase 10.1103/PhysRevLett.110.171601} {\bibfield
  {journal} {\bibinfo  {journal} {Phys.Rev.Lett.}\ }\textbf {\bibinfo {volume}
  {110}},\ \bibinfo {pages} {171601} (\bibinfo {year} {2013})},\ \Eprint
  {http://arxiv.org/abs/1210.2255} {arXiv:1210.2255 [hep-th]} \BibitemShut
  {NoStop}%
\bibitem [{\citenamefont {Bern}\ \emph {et~al.}(2009)\citenamefont {Bern},
  \citenamefont {Carrasco}, \citenamefont {Dixon}, \citenamefont {Johansson},\
  and\ \citenamefont {Roiban}}]{Bern:2009kd}%
  \BibitemOpen
  \bibfield  {author} {\bibinfo {author} {\bibfnamefont {Z.}~\bibnamefont
  {Bern}}, \bibinfo {author} {\bibfnamefont {J.}~\bibnamefont {Carrasco}},
  \bibinfo {author} {\bibfnamefont {L.~J.}\ \bibnamefont {Dixon}}, \bibinfo
  {author} {\bibfnamefont {H.}~\bibnamefont {Johansson}}, \ and\ \bibinfo
  {author} {\bibfnamefont {R.}~\bibnamefont {Roiban}},\ }\href {\doibase
  10.1103/PhysRevLett.103.081301} {\bibfield  {journal} {\bibinfo  {journal}
  {Phys.Rev.Lett.}\ }\textbf {\bibinfo {volume} {103}},\ \bibinfo {pages}
  {081301} (\bibinfo {year} {2009})},\ \Eprint {http://arxiv.org/abs/0905.2326}
  {arXiv:0905.2326 [hep-th]} \BibitemShut {NoStop}%
\bibitem [{\citenamefont {Duff}(1982)}]{Duff:1982yw}%
  \BibitemOpen
  \bibfield  {author} {\bibinfo {author} {\bibfnamefont {M.}~\bibnamefont
  {Duff}},\ }\href@noop {} {\  (\bibinfo {year} {1982})},\ \Eprint
  {http://arxiv.org/abs/1201.0386} {arXiv:1201.0386 [hep-th]} \BibitemShut
  {NoStop}%
\bibitem [{\citenamefont {Bianchi}\ \emph {et~al.}(2008)\citenamefont
  {Bianchi}, \citenamefont {Elvang},\ and\ \citenamefont
  {Freedman}}]{Bianchi:2008pu}%
  \BibitemOpen
  \bibfield  {author} {\bibinfo {author} {\bibfnamefont {M.}~\bibnamefont
  {Bianchi}}, \bibinfo {author} {\bibfnamefont {H.}~\bibnamefont {Elvang}}, \
  and\ \bibinfo {author} {\bibfnamefont {D.~Z.}\ \bibnamefont {Freedman}},\
  }\href {\doibase 10.1088/1126-6708/2008/09/063} {\bibfield  {journal}
  {\bibinfo  {journal} {JHEP}\ }\textbf {\bibinfo {volume} {0809}},\ \bibinfo
  {pages} {063} (\bibinfo {year} {2008})},\ \Eprint
  {http://arxiv.org/abs/0805.0757} {arXiv:0805.0757 [hep-th]} \BibitemShut
  {NoStop}%
\bibitem [{\citenamefont {Anastasiou}\ \emph {et~al.}(2013)\citenamefont
  {Anastasiou}, \citenamefont {Borsten}, \citenamefont {Duff}, \citenamefont
  {Hughes},\ and\ \citenamefont {Nagy}}]{Anastasiou:2013hba}%
  \BibitemOpen
  \bibfield  {author} {\bibinfo {author} {\bibfnamefont {A.}~\bibnamefont
  {Anastasiou}}, \bibinfo {author} {\bibfnamefont {L.}~\bibnamefont {Borsten}},
  \bibinfo {author} {\bibfnamefont {M.}~\bibnamefont {Duff}}, \bibinfo {author}
  {\bibfnamefont {L.}~\bibnamefont {Hughes}}, \ and\ \bibinfo {author}
  {\bibfnamefont {S.}~\bibnamefont {Nagy}},\ }\href@noop {} {\  (\bibinfo
  {year} {2013})},\ \Eprint {http://arxiv.org/abs/1312.6523} {arXiv:1312.6523
  [hep-th]} \BibitemShut {NoStop}%
\bibitem [{\citenamefont {Deser}\ \emph {et~al.}(1982)\citenamefont {Deser},
  \citenamefont {Jackiw},\ and\ \citenamefont {Templeton}}]{Deser:1981wh}%
  \BibitemOpen
  \bibfield  {author} {\bibinfo {author} {\bibfnamefont {S.}~\bibnamefont
  {Deser}}, \bibinfo {author} {\bibfnamefont {R.}~\bibnamefont {Jackiw}}, \
  and\ \bibinfo {author} {\bibfnamefont {S.}~\bibnamefont {Templeton}},\ }\href
  {\doibase 10.1016/0003-4916(82)90164-6} {\bibfield  {journal} {\bibinfo
  {journal} {Annals Phys.}\ }\textbf {\bibinfo {volume} {140}},\ \bibinfo
  {pages} {372} (\bibinfo {year} {1982})}\BibitemShut {NoStop}%
\bibitem [{\citenamefont {Marcus}\ and\ \citenamefont
  {Schwarz}(1983)}]{Marcus:1983hb}%
  \BibitemOpen
  \bibfield  {author} {\bibinfo {author} {\bibfnamefont {N.}~\bibnamefont
  {Marcus}}\ and\ \bibinfo {author} {\bibfnamefont {J.~H.}\ \bibnamefont
  {Schwarz}},\ }\href {\doibase 10.1016/0550-3213(83)90402-9} {\bibfield
  {journal} {\bibinfo  {journal} {Nucl.Phys.}\ }\textbf {\bibinfo {volume}
  {B228}},\ \bibinfo {pages} {145} (\bibinfo {year} {1983})}\BibitemShut
  {NoStop}%
\bibitem [{\citenamefont {de~Wit}\ \emph {et~al.}(1992)\citenamefont {de~Wit},
  \citenamefont {Tollsten},\ and\ \citenamefont {Nicolai}}]{deWit:1992up}%
  \BibitemOpen
  \bibfield  {author} {\bibinfo {author} {\bibfnamefont {B.}~\bibnamefont
  {de~Wit}}, \bibinfo {author} {\bibfnamefont {A.}~\bibnamefont {Tollsten}}, \
  and\ \bibinfo {author} {\bibfnamefont {H.}~\bibnamefont {Nicolai}},\ }\href
  {\doibase 10.1016/0550-3213(93)90195-U} {\bibfield  {journal} {\bibinfo
  {journal} {Nucl.Phys.}\ }\textbf {\bibinfo {volume} {B392}},\ \bibinfo
  {pages} {3} (\bibinfo {year} {1992})},\ \Eprint
  {http://arxiv.org/abs/hep-th/9208074} {hep-th/9208074} \BibitemShut {NoStop}%
\bibitem [{\citenamefont {Witten}(1988)}]{Witten:1988hc}%
  \BibitemOpen
  \bibfield  {author} {\bibinfo {author} {\bibfnamefont {E.}~\bibnamefont
  {Witten}},\ }\href {\doibase 10.1016/0550-3213(88)90143-5} {\bibfield
  {journal} {\bibinfo  {journal} {Nucl.Phys.}\ }\textbf {\bibinfo {volume}
  {B311}},\ \bibinfo {pages} {46} (\bibinfo {year} {1988})}\BibitemShut
  {NoStop}%
\bibitem [{\citenamefont {Ashtekar}\ \emph {et~al.}(1989)\citenamefont
  {Ashtekar}, \citenamefont {Husain}, \citenamefont {Rovelli}, \citenamefont
  {Samuel},\ and\ \citenamefont {Smolin}}]{Ashtekar:1989qd}%
  \BibitemOpen
  \bibfield  {author} {\bibinfo {author} {\bibfnamefont {A.}~\bibnamefont
  {Ashtekar}}, \bibinfo {author} {\bibfnamefont {V.}~\bibnamefont {Husain}},
  \bibinfo {author} {\bibfnamefont {C.}~\bibnamefont {Rovelli}}, \bibinfo
  {author} {\bibfnamefont {J.}~\bibnamefont {Samuel}}, \ and\ \bibinfo {author}
  {\bibfnamefont {L.}~\bibnamefont {Smolin}},\ }\href {\doibase
  10.1088/0264-9381/6/10/001} {\bibfield  {journal} {\bibinfo  {journal}
  {Class.Quant.Grav.}\ }\textbf {\bibinfo {volume} {6}},\ \bibinfo {pages}
  {L185} (\bibinfo {year} {1989})}\BibitemShut {NoStop}%
\bibitem [{\citenamefont {Barton}\ and\ \citenamefont
  {Sudbery}(2003)}]{Barton:2003}%
  \BibitemOpen
  \bibfield  {author} {\bibinfo {author} {\bibfnamefont {C.~H.}\ \bibnamefont
  {Barton}}\ and\ \bibinfo {author} {\bibfnamefont {A.}~\bibnamefont
  {Sudbery}},\ }\href {\doibase 10.1016/S0001-8708(03)00015-X} {\bibfield
  {journal} {\bibinfo  {journal} {Adv. in Math.}\ }\textbf {\bibinfo {volume}
  {180}},\ \bibinfo {pages} {596} (\bibinfo {year} {2003})},\ \Eprint
  {http://arxiv.org/abs/math/0203010} {arXiv:math/0203010} \BibitemShut
  {NoStop}%
\bibitem [{\citenamefont {Baez}(2001)}]{Baez:2001dm}%
  \BibitemOpen
  \bibfield  {author} {\bibinfo {author} {\bibfnamefont {J.~C.}\ \bibnamefont
  {Baez}},\ }\href {\doibase 10.1090/S0273-0979-01-00934-X} {\bibfield
  {journal} {\bibinfo  {journal} {Bull. Amer. Math. Soc.}\ }\textbf {\bibinfo
  {volume} {39}},\ \bibinfo {pages} {145} (\bibinfo {year} {2001})},\ \Eprint
  {http://arxiv.org/abs/math/0105155} {arXiv:math/0105155} \BibitemShut
  {NoStop}%
\bibitem [{\citenamefont {Hodges}(2013)}]{Hodges:2011wm}%
  \BibitemOpen
  \bibfield  {author} {\bibinfo {author} {\bibfnamefont {A.}~\bibnamefont
  {Hodges}},\ }\href {\doibase 10.1007/JHEP07(2013)075} {\bibfield  {journal}
  {\bibinfo  {journal} {Journal of High Energy Physics}\ }\textbf {\bibinfo
  {volume} {1307}} (\bibinfo {year} {2013}),\ 10.1007/JHEP07(2013)075},\
  \Eprint {http://arxiv.org/abs/1108.2227} {arXiv:1108.2227 [hep-th]}
  \BibitemShut {NoStop}%
\bibitem [{\citenamefont {Cachazo}\ \emph {et~al.}(2013)\citenamefont
  {Cachazo}, \citenamefont {He},\ and\ \citenamefont {Yuan}}]{Cachazo:2013iea}%
  \BibitemOpen
  \bibfield  {author} {\bibinfo {author} {\bibfnamefont {F.}~\bibnamefont
  {Cachazo}}, \bibinfo {author} {\bibfnamefont {S.}~\bibnamefont {He}}, \ and\
  \bibinfo {author} {\bibfnamefont {E.~Y.}\ \bibnamefont {Yuan}},\ }\href@noop
  {} {\  (\bibinfo {year} {2013})},\ \Eprint {http://arxiv.org/abs/1309.0885}
  {arXiv:1309.0885 [hep-th]} \BibitemShut {NoStop}%
\bibitem [{\citenamefont {Mason}\ and\ \citenamefont
  {Skinner}(2013)}]{Mason:2013sva}%
  \BibitemOpen
  \bibfield  {author} {\bibinfo {author} {\bibfnamefont {L.}~\bibnamefont
  {Mason}}\ and\ \bibinfo {author} {\bibfnamefont {D.}~\bibnamefont
  {Skinner}},\ }\href@noop {} {\  (\bibinfo {year} {2013})},\ \Eprint
  {http://arxiv.org/abs/1311.2564} {arXiv:1311.2564 [hep-th]} \BibitemShut
  {NoStop}%
\bibitem [{\citenamefont {Landsberg}\ and\ \citenamefont
  {Manivel}(2001)}]{Landsberg2001477}%
  \BibitemOpen
  \bibfield  {author} {\bibinfo {author} {\bibfnamefont {J.}~\bibnamefont
  {Landsberg}}\ and\ \bibinfo {author} {\bibfnamefont {L.}~\bibnamefont
  {Manivel}},\ }\href {\doibase 10.1006/jabr.2000.8697} {\bibfield  {journal}
  {\bibinfo  {journal} {Journal of Algebra}\ }\textbf {\bibinfo {volume}
  {239}},\ \bibinfo {pages} {477 } (\bibinfo {year} {2001})}\BibitemShut
  {NoStop}%
\bibitem [{\citenamefont {Sudbery}(1984)}]{Sudbery:1984}%
  \BibitemOpen
  \bibfield  {author} {\bibinfo {author} {\bibfnamefont {A.}~\bibnamefont
  {Sudbery}},\ }\href {\doibase 10.1088/0305-4470/17/5/018} {\bibfield
  {journal} {\bibinfo  {journal} {J. Phys.}\ }\textbf {\bibinfo {volume}
  {A17}},\ \bibinfo {pages} {939} (\bibinfo {year} {1984})}\BibitemShut
  {NoStop}%
\end{thebibliography}
\end{document}